\documentclass[a4paper,12pt]{article}
\usepackage{epsfig}
\usepackage[dvips,usenames]{color}
\usepackage{graphicx}

\newlength{\dinwidth}
\newlength{\dinmargin}
\def\pslash{\rlap{\hspace{0.02cm}/}{p}}

\begin{document}
\title{Probing the topcolor-assisted technicolor model via the single t-quark production at Hadron colliders}
\bigskip
\author{Xuelei Wang, Yanhui Xi, Yanju Zhang, Huiling Jin\\
 {\small  College of Physics and Information Engineering,}\\
 \small{Henan Normal
University, Xinxiang, Henan, 453007, P.R.China}
\thanks{E-mail:wangxuelei@sina.com}\\}
 \maketitle
\begin{abstract}
\indent The single t-quark production in proton-proton collisions
can proceed through three distinct processes, $tq,~t\bar{b},~tW$
productions. With the running of the Large Hadron Collider(LHC) at
CERN, it has good potential to measure each single t-quark
production mode. The topcolor-assisted technicolor(TC2) model, one
of the promising dynamical theories, predicts some new particles
at several hundred GeV scale: three top-pion
bosons($\Pi^{\pm}_{t},\Pi^{0}_{t}$) and one top-higgs
boson($h_{t}$). These particles are regarded as the typical
feature of the TC2 model and can contribute to some processes. In
this paper, we systematically study the contribution of the TC2
model to the single t-quark production at the Hadron colliders,
specially at the LHC. The TC2 model can contribute to the cross
section of the single t-quark production in two different ways.
First, the existence of the top-pions and top-higgs can modify the
$Wtb$ coupling via their loop contributions, and such modification
can cause the correction to the cross sections of all three
production modes. Our study shows that this kind of correction is
negative and very small in all cases. Thus it is difficult to
observe such correction even at the LHC. On the other hand, there
exist the tree-level FC couplings in the TC2 model which can also
contribute to the cross sections of the $tq$ and $t\bar{b}$
production processes. The resonant effect can greatly enhance the
cross sections of the $tq$ and $t\bar{b}$ productions. The first
evidence of the single t-quark production has been reported by the
$D0$ collaboration and the measured cross section for the single
t-quark production of $\sigma(p\bar{p}\rightarrow tb+X,tqb+X)$ is
compatible at the $10\%$ level with the standard model prediction.
Because the light top-pion can make great contribution to the
$t\bar{b}$ production, the top-pion mass should be very large in
order to make the predicted cross section in the TC2 model be
consistent with the Tevatron experiments. More detailed
information about the top-pion mass and the FC couplings in the
TC2 model should be obtained with the running of the LHC.
\end{abstract}

Pacs: 12.60Nz,14.80.Mz,12.15.Lk

\newpage
\section {Introduction}
 \indent
The t-quark, as the most massive object in the Standard Model(SM),
is that which felt the symmetry breaking the most profoundly. The
t-quark physics has become a very active research area since the
existence of the t-quark was established in t-quark pair events
produced via the strong interaction\cite{top pair}, where
quark-antiquark annihilation or gluon-gluon fusion leads to
top-antitop pairs. In addition to the top pair production via the
strong interaction at hadron colliders, they can also be produced
singly in electroweak(EW) interactions, and there are three single
t-quark production modes in the SM: The t-channel tq production,
the s-channel $t\bar{b}$ production and the associated $tW$
production. Studying these single t-quark production modes at
hadron colliders is important for a number of reasons. First, a
measurement of the production cross section provides the only
direct measurement of the total t-quark decay width and the CKM
matrix element $|V_{tb}|^2$, without having to assume three quark
generations or CKM matrix unitarity. Second, the measurement of
the spin polarization of single t-quarks can be used to test the
V-A structure of the t-quark EW charged current interaction.
Third, the presence of various new SM and non-SM phenomena may be
inferred by observing deviations from the predicted rate of the
single t-quark signal and by comparing different production modes.
Fourth, the single t-quark final states present an irreducible
background to several searches for SM or non-SM signals, for
example Higgs boson searches in the associated production channel.

Here, we should emphasize that the single t-quark production is
interesting beyond the SM. The single t-quark production as a
window to probe new physics has been systematically
studied\cite{new phys-3,new phys-0} and the study shows that three
single top production modes can be used to distinguish several new
physics models. New physics can influence the single t-quark
production by inducing non-standard weak interactions\cite{new
phys-0,new phys-1,new phys-2}, via loop effects\cite{new
phys-4,new phys-5}, or by providing new sources of single t-quark
events\cite{new phys-0,new phys-2,new phys-5,new phys-6}. Three
single t-quark production modes respond quite differently to
different realizations of physics beyond the SM\cite{new phys-0}.
In general, the $tq$ production mode is insensitive to heavy
charged bosons. The reason for this is that the t-channel
exchanged results in a space-like momentum, which never can go
on-shell, and thus the amplitude for the heavy particle is always
suppressed by the mass of the heavy boson, $1/M^2_B$. However, the
FCNC processes can have a drastic effect on the $tq$ production
mode. Because they involve new FC interactions between the t-quark
and a light quark(c or u), the $tq$ production mode can be
enhanced significantly. The $t\bar{b}$ production mode is very
sensitive to an exotic charged boson which couples to t-quark and
b-quark. Because the exchanged particle is time-like, there is the
possibility(if it is heavier than the t-quark) that it can be
produced on-shell, resulting in a large enhancement of the cross
section. Specific theories which predict an enhancement of the
cross section of the $t\bar{b}$ production are theories with a
$W'$\cite{new phys-7} or charged Higgs, both of which can result
in the cross section of the $t\bar{b}$ production different from
the SM by factors of few at either the Tevatron or the
LHC\cite{new phys-0,new phys-8}. For the $tW$ mode, the cross
section is more or less insensitive to new bosons, because the $W$
is manifest in the final state. Furthermore, the $tq$ and
$t\bar{b}$ modes are sensitive to different new physics models and
hence can be used to distinguish between various exotic models.
From this line of thinking, we see that all three modes are really
complimentary views of the t-quark and new physics, and thus
measured separately they provide more information than would be
obtained by lumping them together into a singular single-top
process.

On the experimental aspect, twelve years after the discovery of
the t-quark via strong pair production at the Tevatron, the first
evidence of the single t-quark production has been reported by the
$D0$ collaboration\cite{single top quark1,single top quark2} at
the Tevatron. The events were selected from a $0.9 fb^{-1}$
dataset that have an electron or muon and missing transverse
energy from the decay of a $W$ boson from the top quark decay, and
two, three, or four jets with one or two of the jets identified as
originating from a b hadron decay. A binned likelihood fit of the
signal cross section plus background to the data from the
combination of the results from the three analysis methods gives a
cross section for single t-quark production of
$\sigma(p\bar{p}\rightarrow tb+X,tqb+X)=4.7\pm 1.3 pb$\cite{single
top quark1}. Such value is compatible at the $10\%$ level with the
standard model prediction. The LHC will accelerate proton beams
and bring them to collision at a center of mass (c.m.) energy of
$\sqrt{s}=14$ TeV and at luminosities between $(1-2)\times
10^{33}cm^{-2}s^{-1}$(initial low-luminosity phase) and
$10^{34}cm^{-2}s^{-1}$(high luminosity). The single t-quark
production can be discovered at the Tevatron but the Tevatron has
little ability to observe three modes individually. The study of
the single t-quark production is a very important part of research
programs at future LHC experiments. Such a study allows to
investigate the t-quark properties with high enough accuracy and
to measure a $Wtb$ coupling structure with high precision. It may
shed a light on the underlying theory which probably stands beyond
the SM. At the LHC, it is expected that three different single
t-quark production modes can be observed individually. The three
single t-quark processes result in quite distinct final states and
topologies, leading to the definition of specific analyses in each
case. The discrimination between them makes use of difference in
jet multiplicity, number of b-tagged jets required. Besides,
important difference subsists in the level of backgrounds that are
faced in the various analyses, leading to the development of tools
dedicated to the rejection of specific backgrounds. With more than
two millions single t-quark events produced every year during a
low luminosity run at the LHC, a precise determination of all
contributions to the total single t-quark cross section seems
achievable. These measurements will constitute the first direct
measurement of $V_{tb}$ at the few percent level of precision, and
also constitute a powerful probe for new physics, via the search
for evidence of anomalous couplings to the t-quark, or the
measurement of additional bosonic contributions to the single
t-quark production.

In the transition from the Tevatron to the LHC, several aspects of
single t-quark physics change. At the Tevatron, the main goal is
to observe the EW mode of t-quark production for the first time,
and that will be followed by initial measurements. Hence, the
emphasis is on extracting the signal from the backgrounds, using
optimized methods. By contrast, by the time that the LHC analyses
are starting, the single t-quark production should already have
been discovered, and the focus shifts to precision measurement.
Thus, the single t-quark production can be used as tools to probe
the EW sector and to look for new physics.

Among the various new physics models, the topcolor-assisted
technicolor (TC2) model\cite{Hill,Lane,Cvetic,Buchalla} is a most
promising candidate of dynamical theories. The TC2 model gives a
reasonable explanation of the electroweak symmetry breaking(EWSB)
and heavy t-quark mass. In the TC2 model, the topcolor interaction
makes small contribution to the EWSB, and gives rise to the main
part of the t-quark mass $(1-\varepsilon)m_{t}$ with a model
dependant parameter $0.03\leq\varepsilon\leq 0.1$\cite{Buchalla}.
The technicolor interaction plays a
 main role in the breaking of EW gauge symmetry. To
 account for the explicit breaking of quark and lepton flavor
 symmetries, the extended technicolor(ETC) was invented.
 The ETC interaction gives rise to
 the masses of the ordinary fermions including a very small portion
 of the t-quark mass $\varepsilon m_{t}$. This kind of model predicts
 three CP odd top-pions $(\Pi^{0}_{t}$, $\Pi^{\pm}_{t})$ and one CP even top-higgs($h_t^0$) with large
 Yukawa couplings to the third family. As we know, the topcolor
interaction is non-universal and such non-universal feature can
result in the new tree-level flavor-changing(FC) couplings when
one writes the interaction in the quark mass eigen-basis. The TC2
model can contribute to the single t-quark production in two
different ways. One is that the top-pions and top-higgs can make
the loop contribution to the vertex $Wtb$, and such contribution
can influence the production rates of all three single t-quark
production modes. Another is that the existence of the FC
couplings in the TC2 model can make a significant tree-level
contribution to the FC t-quark production processes and some
studies have been done\cite{FC-TC2}. The running of the LHC will
open an ideal window to probe the effect of the TC2 model via the
single t-quark production. In this paper, we will calculate these
two kinds of contributions and study the potential to probe the
TC2 model via the single t-quark production at hadron colliders,
specially at the LHC.

  The rest of this paper is organized as follows. In section 2, we will give a detail calculation of the contributions
of the TC2 model to each single t-quark production mode at hadron
colliders and discuss the numerical results.  The summary and
conclusion are represented in section 4.

\section{The contribution of the TC2 model to each single t-quark
production mode at Hadron colliders}

\subsection{One-loop correction of the TC2 model to the $Wtb$ coupling}

As it is known, the top-pions and top-higgs are predicted by the
TC2 model. The couplings of the top-pions and top-higgs to the
three family fermions are non-universal, and the top-pions and
top-higgs have large Yukawa couplings to the third family. Such
feature can result in the significant tree-level FC couplings of
the top-pions and top-higgs to the quarks when one writes the
interaction in the quark mass eigen-basis. The couplings of the
top-pions and top-higgs to a pair of quarks are proportion to the
masses of the quarks and the explicit form can be written as
\cite{He}

 \begin{eqnarray}
 \mathcal
{L}&=&\frac{m_{t}}{v_{w}}\tan\beta[iK^{tt}_{UR}K^{tt*}_{UL}\overline{t_{L}}t_{R}\Pi^{0}_{t}
+\sqrt{2}K^{tt*}_{UR}K^{bb}_{DL}\overline{t_{R}}b_{L}\Pi^{+}_{t}+\sqrt{2}K^{tc*}_{UR}K^{bb}_{DL}\overline{c_{R}}b_{L}\Pi^{+}_{t}\\
 \nonumber
 &&
+i\frac{m^{*}_{b}}{m_{t}}\overline{b_{L}}b_{R}\Pi^{0}_{t}+K^{tt}_{UR}K^{tt*}_{UL}\overline{t_{L}}t_{R}h^{0}_{t}
+iK^{tc}_{UR}K^{tt*}_{UL}\overline{t_{L}}c_{R}\Pi^{0}_{t}+K^{tc}_{UR}K^{tt*}_{UL}\overline{t_{L}}c_{R}h^{0}_{t}+h.c.].
 \end{eqnarray}
 Where
 $\tan\beta=\sqrt{(v_{w}/v_{t})^{2}-1}$, $v_{t}=60-100 $ GeV is the
 top-pion decay constant, $v_{w}=246$ GeV is the EWSB
 scale, $K^{i,j}_{U,D}$ are the matrix elements of the unitary
 matrix $K_{U,D}$, from which the Cabibbo-Kobayashi-Maskawa (CKM)
 matrix can be derived as $V=K^{-1}_{UL}K_{DL}$. Their values can
 be written as

  \begin{eqnarray*}
K^{tt}_{UL}=K^{bb}_{DL}\approx1, ~~~~ ~~
K^{tt}_{UR}=1-\varepsilon, ~~~~~~
K^{tc}_{UR}=\sqrt{2\varepsilon-\varepsilon^{2}}.
 \end{eqnarray*}
 Here $\varepsilon=0.03-0.1$, the mass $m^{*}_{b}$ is a part of b-quark mass which is induced
 by the instanton, and can be estimated as\cite{Hill, Lane}
  \begin{eqnarray*}
  m^{*}_{b}=\frac{3m_{t}\kappa}{8\pi^{2}}\sim6.6\kappa~~ GeV,
\end{eqnarray*}
which we generally expect $\kappa\sim1$ to $10^{-1}$ as in QCD.

The existence of the top-pions and top-higgs can make the loop
contribution to the $Wtb$ coupling. The leading order(LO)
contribution arises from the terms including t-quark mass $m_t$.
Considering the LO contribution, we only need to calculate the
diagrams shown in Fig.1.

The renormalized effective $Wtb$ coupling can be written as
 \begin{eqnarray}
    \Gamma^{\mu}(p_t,p_b)&=&-i\frac{g}{\sqrt[]{2}}\{\gamma^{\mu}P_{L}[1+F_{L}+\frac{1}{2}\delta\mathbf{Z}^{L}_{b}
 +\frac{1}{2}\delta\mathbf{Z}^{L}_{t}]+\gamma^{\mu}P_{R}F_{R}\\ \nonumber
 & &+p^{\mu}_{t}[P_{L}\widetilde{F}_{L}+P_{R}\widetilde{F}_{R}]
 +p^{\mu}_{b}[P_{L}\widehat{F}_{L}+P_{R}\widehat{F}_{R}]\}.
\end{eqnarray}
 Here
 $P_{L}=\frac{1}{2}(1-\gamma_{5})$ and $P_{R}=\frac{1}{2}(1+\gamma_{5})$
 are the chirality projectors. $p^{\mu}_{t}$ and $p^{\mu}_{b}$ are the momenta of outgoing t-quark and incoming b-quark.
 The form factors
 $F_{L,R}$ and $\widetilde{F}_{L,R}$ and $\widehat{F}_{L,R}$
 represent the contributions from the irreducible vertex
 loops. $\delta\mathbf{Z}^{L}_{b}$ and $\delta\mathbf{Z}^{L}_{t}$
 denote the field renormalization constants for
 b-quark and t-quark, respectively.
 In the calculation of the renormalized effective $Wtb$ coupling, we take dimensional regularization and on-shell
renormalization scheme. The explicit expressions of the form
factors are given by(we have neglected b-quark mass)
\begin{eqnarray}
    F_{L}&=&\frac{1}{16\pi^{2}}\frac{m^{2}_{t}\tan^{2}\beta}{v^{2}_{w}}(1-\varepsilon)^{2}(2C^{e}_{24}+2C^{f}_{24}),\\
    F_{R}&=&0,\\
 \delta\mathbf{Z}^{L}_{b}&=&\frac{1}{8\pi^{2}}\frac{m^{2}_{t}\tan^{2}\beta}{v^{2}_{w}}(1-\varepsilon)^{2}B^{d}_{1},\\
 \delta\mathbf{Z}^{L}_{t}&=&\frac{1}{16\pi^{2}}\frac{m^{2}_{t}\tan^{2}\beta}{v^{2}_{w}}(1-\varepsilon)^{2}[B^{b}_{1}+B^{c}_{1}
  +2m^{2}_{t}(B^{b'}_{1}+B^{c'}_{1}+B^{a'}_{1}+B^{b'}_{0}+B^{c'}_{0})],\\
   \widetilde{F}_{L}&=&\frac{1}{16\pi^{2}}\frac{m^{3}_{t}\tan^{2}\beta}{v^{2}_{w}}(1-\varepsilon)^{2}(3C^{e}_{11}+
   C^{e}_{0}+2C^{e}_{21}+2C^{e}_{22}-3C^{e}_{12}-4C^{e}_{23}
   +C^{f}_{12}\\ \nonumber
   &&
   -C^{f}_{11}-C^{f}_{0}
   +2C^{f}_{21}+2C^{f}_{22}-4C^{f}_{23}),\\
   \widetilde{F}_{R}&=&0,\\
   \widehat{F}_{L}&=&\frac{1}{16\pi^{2}}\frac{m^{3}_{t}\tan^{2}\beta}{v^{2}_{w}}(1-\varepsilon)^{2}(C^{e}_{12}+C^{e}_{11}+C^{e}_{0}
   -2C^{e}_{22}+2C^{e}_{23}-3C^{f}_{12}+C^{f}_{11}-C^{f}_{0}\\
\nonumber
   &&-2C^{f}_{22}+2C^{f}_{23}),\\
   \widehat{F}_{R}&=&0.
 \end{eqnarray}
 $B_{0,1}$ and $C_{0,ij}$ are respectively
 two-point and three-point standard functions given in reference\cite{standard
 functions} and $B'_{0,1}$ denotes
$\partial$$B_{0,1}$/$\partial$$p^{2}$. Their functional
dependences are

 \begin{eqnarray*}
  C^{e}_{0,ij}&=&C_{0,ij}(-p_{t},p_{t}-p_{b},m_{t},M_{\Pi_{t}},M_{\Pi_{t}}),\\
  C^{f}_{0,ij}&=&C_{0,ij}(-p_{t},p_{t}-p_{b},m_{t},M_{h_{t}},M_{\Pi_{t}}),\\
  B^{a}_{0,1}&=&B_{0,1}(-p_{t},m_{b},M_{\Pi_{t}}),\\
  B^{b}_{0,1}&=&B_{0,1}(-p_{t},m_{t},M_{\Pi_{t}}),\\
  B^{c}_{0,1}&=&B_{0,1}(-p_{t},m_{t},M_{h_{t}}),\\
  B^{d}_{0,1}&=&B_{0,1}(-p_{b},m_{t},m_{\Pi_{t}}).
   \end{eqnarray*}
  Here we have ignored the mass difference between the neutral
  top-pion and charged top-pions.

 The correction of the TC2 model to
 the $Wtb$ coupling can influence all three single t-quark
 production modes.

\subsection{The tq production mode}
 In the SM, the tree-level
$tq$ production mode($q=\bar{u},d,\bar{c},s$, here q also
represent anti-quarks for simplicity. In the following, we will
name this process as $tq$ production mode instead of t-channel
mode called in the SM because the tree-level FC couplings in the
TC2 model can also make a s-channel contribution to the $tq$
production.) involves a spacelike $W$ boson $(Q^2\leq 0)$, as
shown in Fig.2(a), and the virtual $W$ boson strikes a b-quark in
the proton sea, promoting it to a t-quark. Of the single t-quark
production modes, the $tq$ production mode has the largest cross
section both at the Tevatron and the LHC and such mode has been
studied extensively\cite{t cross section}. Its cross section has
been calculated at next-to-leading-order(NLO) in QCD, and the QCD
NLO total cross section is 1.98 pb, 246 pb for the Tevatron and
the LHC, respectively\cite{t-theoretical
uncertainty-1,t-theoretical uncertainty-2,t-theoretical
uncertainty-3}. Significant sources of uncertainties affect the
theoretical predictions of the cross section and the theoretical
uncertainty of the $tq$ production cross section is the largest
one in all single t-quark production modes. At the LHC, the cross
section is so large that it should be possible to collect large
samples of single t-quark events via the $tq$ production which can
be used to study the t-quark EW coupling in detail.

The modification of the $Wtb$ coupling can influence the cross
section of the $tq$ production in the way shown in Fig.3.

The corresponding production amplitudes can be written as
\begin{eqnarray}
 M_{1}=-\frac{g}{\sqrt{2}}G(p_{4}-p_{2},M_{W})\overline{u}_{t}(p_{4})
 \Gamma^{\mu}(p_4,p_2)u_{b}(p_{2})\overline{u}_{q'}(p_{3})\gamma_{\mu}P_{L}u_{q}(p_{1}),
 \end{eqnarray}
\begin{eqnarray}
 M_{2}=-\frac{g}{\sqrt{2}}G(p_{4}-p_{2},M_{W})\overline{u}_{t}(p_{4})
 \Gamma^{\mu}(p_4,p_2)u_{b}(p_{2})\overline{v}_{\overline{q}'}(p_{1})\gamma_{\mu}P_{L}v_{\overline{q}}(p_{3}),
 \end{eqnarray}
 where
\begin{eqnarray*}
  G(p,m)=\frac{1}{p^{2}-m^{2}}.
 \end{eqnarray*}

 On the other hand, as we can see from equation (1), there exist
 the tree-level FC couplings
 $\Pi_t^0t\bar{c},~h_t^0t\bar{c},~\Pi_t^+c\bar{b}$ in the TC2 model. Via $b\bar{b}$ collision, these FC
 couplings can induce the tree-level contribution to the $t\bar{c}$ production via the t-channel and
 s-channel exchange of $\Pi_t^0,h_t^0,\Pi_t^+$. The corresponding
 Feynman diagrams are shown in Fig.4. We should note that the cross section of the
  $t\bar{c}$ production via the t-channel in the SM
is very small and can be ignored. The tree-level FC contribution
of the TC2 model, specially the resonant effect in the s-channel
 of Fig.4, can greatly enhance the cross section of
the $t\bar{c}$ production. The FC contribution of the TC2 model
does not have interference with the SM amplitudes, because the FC
contribution only comes from $b\bar{b}$ collision. Although such
FC contribution arises from both t-channel and s-channel, we
combine all the cross sections of $tq(q=\bar{u},d,\bar{c},s)$
productions together due to the difficulty to distinguish these
light quarks.

The production amplitudes related to Fig.4 can be written as
\begin{eqnarray}
 M_{3}&=&-i\frac{2m^{2}_{t}\tan^{2}\beta}{v^{2}_{w}}(1-\varepsilon)\sqrt{2
 \varepsilon-\varepsilon^{2}}\frac{1}{(p_{4}-p_{2})^{2}-M^{2}_{\Pi_{t}}}
 \overline{u}_{t}(p_{4})P_{L}u_{b}(p_{2})
 \overline{v}_{\overline{b}}(p_{1})P_{R}v_{\overline{c}}(p_{3}),\\
 M_{4}&=&i\frac{m_{t}m^{*}_{b}\tan^{2}\beta}{v^{2}_{w}}\sqrt{2
 \varepsilon-\varepsilon^{2}}\frac{1}{(p_{3}+p_{4})^{2}-M^{2}_{\Pi_{t}}
 +iM_{\Pi_{t}}\Gamma_{\Pi^{0}_{t}}}\overline{u}_{t}(p_{4})P_{R}v_{\overline{c}}(p_{3})
 \overline{v}_{\overline{b}}(p_{1})\gamma_{5}u_{b}(p_{2}),\\
 M_{5}&=&-i\frac{m_{t}m^{*}_{b}\tan^{2}\beta}{v^{2}_{w}}\sqrt{2
 \varepsilon-\varepsilon^{2}}\frac{1}{(p_{3}+p_{4})^{2}-M^{2}_{h^{0}_{t}}
 +iM_{h^{0}_{t}}\Gamma_{h^{0}_{t}}}\overline{u}_{t}(p_{4})P_{R}v_{\overline{c}}(p_{3})
 \overline{v}_{\overline{b}}(p_{1})u_{b}(p_{2}).
 \end{eqnarray}

We can see that the time-like momentum may hit the
top-pion(top-higgs) pole in the top-pion(top-higgs) propagator of
s-channel in Fig.4. So we should take into account the effect of
the width of the top-pions(top-higgs) in the amplitudes $M_4,M_5$.
i.e., we should take the complex mass term $M^{2}_{\Pi_{t}}
-iM_{\Pi_{t}}\Gamma_{\Pi^{0}_{t}}(M^{2}_{h_{t}}
-iM_{h_t}\Gamma_{h^{0}_{t}})$ instead of the simple
top-pions(top-higgs) mass term $M^{2}_{\Pi_{t}}(M^{2}_{h_{t}})$ in
the top-pions(top-higgs) propagator. The  term $M^{2}_{\Pi_{t}}
-iM_{\Pi_{t}}\Gamma_{\Pi^{0}_{t}}(M^{2}_{h_{t}}
-iM_{h_t}\Gamma_{h^{0}_{t}})$ is important in the vicinity of the
resonance. The decay widths of $\Pi_t^0$ and $h_t^0$ have been
calculated in reference\cite{decay widths}.

With the above production amplitudes, we can directly obtain the
cross sections of subprocess. The hadronic cross sections at the
Hadron colliders can be obtained by folding the cross sections of
subprocesses with the parton distribution. In the calculation of
the cross section, instead of calculating the square of the
production amplitudes analytically, we calculate the amplitudes
numerically by using the method of reference\cite{HZ}. This
greatly simplifies our calculations.

 To obtain numerical results of the contribution of the TC2 model to
  the $tq$ production, we need to specify the relevant
parameters in the SM. These parameters are $m_{t}=$174.2 GeV,
$m_{b}=$4.7 GeV, $m_{c}=$1.25 GeV, $m_{u}=$0.002 GeV,
$m_{d}=$0.005 GeV, $m_{s}=$0.095 GeV, $\alpha_{s}=$0.118,
$s^{2}_{W}=$0.23, and $M_{W}=$80.4 GeV\cite{parameters}. Expect
for these parameters, the production amplitudes are also dependent
on some free parameters in the TC2 model: $v_t$, $\varepsilon$,
$M_{\Pi_t}$, $M_{h_t}$. Here we choose $v_t=60$ GeV and $tan\beta$
can also be fixed. To see the influences of these parameters on
the cross sections, we take $\varepsilon=0.03, 0.06, 0.1$ and
$M_{h_t}=200, 400$ GeV as examples, and vary $M_{\Pi_t}$ from 200
GeV to 400 GeV. For the parton distributions, we use CTEQ6L
PDF\cite{PDF}. On the other hand, the conjugated $\bar{t}$
production is also considered in the calculation.

To see the effect of the varying $M_{\Pi_t}$, in Fig.5, we plot
the relative correction
$(\sigma^{tq}_{total}-\sigma^{tq}_{SM})/\sigma^{tq}_{SM}$ at the
LHC as a function of $M_{\Pi_t}$ for three values of
$\varepsilon(0.03,0.06,0.1)$ and two values of $M_{h_t}$(200,400
GeV). $\sigma^{tq}_{total}$ is the total cross section of $tq$
production at the LHC which is defined as
$\sigma^{tq}_{total}=\sigma^{tq}_{SM}+\delta\sigma^{tq}_{Wtb}+\sigma^{tq}_{FC}$,
with $\sigma^{tq}_{SM}$ being the tree-level SM cross section of
the $tq$ production and $\delta\sigma^{tq}_{Wtb}$ being the
correction to the cross section induced by the modification of the
$Wtb$ coupling in the TC2 model and $\sigma^{tq}_{FC}$ being the
cross section of the $t\bar{c}$ production induced by the
tree-level FC couplings in the TC2 model. From the Fig.5, we can
see that the total correction of the TC2 model is positive in most
case and it becomes negative when $M_{\Pi_t}$ is large. This is
because $\delta\sigma^{tq}_{Wtb}$ is negative and
$\sigma^{tq}_{FC}$ is positive. The resonant effect of s-channel
enhances the $\sigma^{tq}_{FC}$ significantly and it drops sharply
when $M_{\Pi_t}$ becomes large. As we will discuss in the next
section, the light top-pion is not allowed by the Tevatron
experiments due to the large contribution of the light top-pion to
the $t\bar{b}$ production. For the heavy top-pion, the
contribution of the TC2 model to the $tq$ production at the LHC
are not large. Comparing two diagrams in Fig.5, we can conclude
that large $M_{h_t}$ can also depress the correction. The
dependence of $\varepsilon$ on the relative correction is clear,
and the relative correction increases with $\varepsilon$
increasing.

 To provide more information,  we show each contribution of the
TC2 model to the $tq$ production at the LHC in Table 1. We can see
that the main correction of the TC2 model to the $tq$ production
comes from the tree-level FC coupling and such correction is
positive. The correction induced by the modification of the $Wtb$
coupling in the TC2 model is negative and the maximal value of the
relative correction is only about $-4\%$.

The ability to detect the correction of the TC2 model via the
single t-quark production at the LHC is determined by the
precision to measure the $tq$ process. The precision on the cross
section is related to the statistical sensitivity, systematic
uncertainties, theoretical background uncertainties and the
luminosity uncertainties. So the final state signature and the
backgrounds should be analyzed in detail.

The final state signature of the $tq$ production is characterized
by a high energy isolated lepton and missing transverse energy
from the decay of the W from the t-quark into $l\nu$, and two or
three jets. One of the jets originates from a b-quark from the
t-quark decay and is usually central(low pseudorapidities) and
energetic. There usually are, apart from the b jet from t-quark
decay, a moderately energetic light flavor jet and a high
pseudorapidity low energy b-quark jet from gluon splitting. This
very forward or backward b jet is a unique feature of this signal,
but it is rarely reconstructed and even more difficult to tag.
Among the two or three jets, at least one jet must be b tagged in
the central pseudo-rapidity region. The other b jet in the final
state is usually emitted towards the very forward region, outside
the tracker acceptance and thus out of reach of the b-tagging
algorithm in most case.

The main processes that can mimic the final state topology of $tq$
production are: (i)W+jets events, where the W boson decays
semileptonically and two or more associated jets are produced;
(ii)$t\bar{t}$ events, where one or both t-quarks decay
leptonically; and (iii)QCD or multi-jet events. Contrary to the
situation at the Tevatron, the main background comes from the
t-quark pair production at the LHC, well above the W+jets and WQQ
events. The t-quark pair production has a cross section larger
than the single t-quark production. But the average energy in the
event is larger, due to the presence of two t-quarks, and events
tend to be more spherical and have more jet multiplicity than
single t-quark events. Two t-quarks produce two W bosons and two
b-quark jets, the latter with very similar kinematics to the
signal and therefore likely to be b tagged as well. The same final
state signature as in the single t-quark processes is obtained if
only one of the W bosons decays leptonically  and the other
hadronically, or if both do, but only one lepton is reconstructed.
This background can be properly simulated using ALPGEN or PYTHIA.
The W+jets background is by far the most problematic to get rid
of. It consists of a leptonically decaying W boson and at least
two associated quarks or gluons. W+jets events contain less energy
in the event than the single t-quark signals since they do not
contain a heavy object like the t-quark. But the cross section is
very large in comparison to the single t-quark production, and the
flavor composition of the associated jets is sufficiently complex,
to make this background hard to model and even harder to get rid
of as one applies b tagging techniques, since they tend to shift
distributions to be more signal-like and wash away any low energy
feature. This background has been estimated using simulated
events, by ALPGEN for example, and is usually scaled to data to
get the overall normalization right.  The QCD background typically
enters as misreconstructed events, where a jet is wrongly
identified as an electron, or a muon from a heavy flavor jet
appears isolated in the detector. Multi-jet events may also
contain heavy flavor jets or just light jets that are
misidentified by the b tagging algorithma. The transverse energy
of QCD events is much less than signal events, and the mass of the
system of the b-tagged jet, the lepton and the neutrino does not
peak at $m_t$, but the cross section is overwhelmingly large. This
background is usually obtained directly from data, and after some
initial basic criteria can be reduced in size to the same level as
the signal.

With a large cross section, the $tq$ production will be the first
single t-quark production accessible with the early data at the
LHC. The cross section measurement of such mode benefits from a
significantly higher statistics compared to the other single
t-quark production modes. The final topology is also significantly
different from that of the other modes, and leads to a specific
selection. The precision on the cross section is related to the
statistical sensitivity, systematic uncertainties, theoretical
background uncertainties and the luminosity uncertainties.  With a
simple selection, the precision is expected to be\cite{ATLAS
precision}
\begin{eqnarray}
\frac{\Delta \sigma}{\sigma}=1.02\%_{stat}\pm 11\%_{exp}\pm
6\%_{bckgd theo}\pm 5\%_{lumi}
 \end{eqnarray}
 at ATLAS for $L=30 fb^{-1}$, and
\begin{eqnarray}
\frac{\Delta \sigma}{\sigma}=2.7\%_{stat}\pm 8.1\%_{exp}\pm
3\%_{lumi}
 \end{eqnarray}
at CMS for $L=10 fb^{-1}$\cite{CMS precision}.

We can see that the statistical sensitivity is very small with the
large cross section at the LHC and the main uncertainties come
from systematic uncertainties. Based on our calculation, we can
conclude that the LHC should have the ability to detect the
contribution induced by the FC coupling in the TC2 model via the
$tq$ production, but higher sensitivity is needed if one wants to
obtain the information about the modification of the $Wtb$
coupling in the TC2 model.

 \subsection{The $t\bar{b}$ production mode}
Another process in the SM that produces a single t-quark is the
s-channel $t\bar{b}$ production via the time-like W boson, as
shown in Fig.2(b)\cite{s cross section}. The cross section of
 such production in the SM is much less than that of the
$tq$ production because it scales like $1/s$ rather than
$1/M^2_W$. However, the $t\bar{b}$ process has the advantage of
little theoretical uncertainty\cite{t-theoretical
uncertainty-2,t-theoretical uncertainty-3,s-theoretical
uncertainty}. This is because the quark and anti-quark
distribution functions are relative well known, so the uncertainty
from the parton distribution functions is small. Furthermore, the
parton luminosity can be constrained by measuring the Drell-Yan
process $q\bar{q}\rightarrow W^*\rightarrow l\bar{\nu}$, which has
the identical initial state. The NLO cross section of the
$t\bar{b}$ production in the SM is 0.88 pb, 10.6 pb for the
Tevatron and the LHC, respectively\cite{t-theoretical
uncertainty-2,t-theoretical uncertainty-3,s-theoretical
uncertainty}. The total cross section is even known to
next-to-next-to-leading order\cite{s-NNLO}.

Due to the existence of the $Wtb$ coupling, the modification of
the $Wtb$ coupling in the TC2 model also affects the $t\bar{b}$
production as shown in Fig.6(6).

Including the modification of the $Wtb$ coupling in the TC2 model,
we can write the production amplitude of the $t\bar{b}$ process as
\begin{eqnarray}
 M_{6}=-\frac{g}{\sqrt[]{2}}G(p_{3}+p_{4},M_{W})\overline{u}_{t}(p_{4})
 \Gamma^{\mu}(p_4,-p_3)v_{\overline{b}}(p_{3})\overline{v}_{\overline{q}'}(p_{2})\gamma_{\mu}P_{L}u_{q}(p_{1}).
 \end{eqnarray}

On the other hand, due to the existence of the tree-level FC
coupling $\Pi^+c\bar{b}$, $\Pi_t^{+}$ can contribute to the
$t\bar{b}$ production through the s-channel virtual exchange of a
$\Pi_t^{+}$ as shown in Fig.6(7).

As in the $tq$ production, the s-channel contribution of
$\Pi_t^{+}$ to the $t\bar{b}$ production can also allow large
resonant contribution. The distribution of the invariant mass of
the $t\bar{b}$ system could show the resonant effect around the
mass of $\Pi_t^{+}$, which serves to identify this type of
particles. However, if the mass of $\Pi_t^{+}$ is very large and
its width is broad, the resonant shape can be washed out.

The production amplitude of the $t\bar{b}$ process induced by the
FC coupling $\Pi^+c\bar{b}$ is
 \begin{eqnarray}
 M_{7}&=&-i\frac{2m^{2}_{t}\tan^{2}\beta}{v^{2}_{w}}(1-\varepsilon)\sqrt{2
 \varepsilon-\varepsilon^{2}}\frac{1}{(p_{3}+p_{4})^{2}-M^{2}_{\Pi_{t}}
 +iM_{\Pi_{t}}\Gamma_{\Pi^{+}_{t}}}\overline{u}_{t}(p_{4})P_{L}v_{\overline{b}}(p_{3})\\
\nonumber
 &&\overline{v}_{\overline{b}}(p_{1})P_{R}u_{c}(p_{2}).
 \end{eqnarray}
Here we also take into account the effect of the width of the
charged top-pions in the amplitude $M_7$ due to the existence of
the resonant effect. The decay width of the charged top-pions has
been given in reference\cite{decay width-charged}

The amplitude $M_7$ does not have a significant interference with
the SM amplitudes because the SM contribution is mostly from light
quarks($u$ and $\bar{d}$). So, in the TC2 model, the total cross
section of the $t\bar{b}$ production can been written as

\begin{eqnarray}
 \sigma^{t\bar{b}}_{total}=\sigma^{t\bar{b}}_{SM}+\delta\sigma^{t\bar{b}}_{Wtb}+\sigma^{t\bar{b}}_{FC}.
 \end{eqnarray}
Here $\sigma^{t\bar{b}}_{SM}$ is the SM tree-level cross section
of $t\bar{b}$ production, $\delta\sigma^{t\bar{b}}_{Wtb}$ is the
correction induced by the modification of the $Wtb$ coupling in
the TC2 model, and $\sigma^{t\bar{b}}_{FC}$ is the cross section
of the $t\bar{b}$ production induced by the tree-level FC coupling
in the TC2 model.

In Fig.7, we plot the total relative correction at the LHC,
$(\sigma^{t\bar{b}}_{total}-\sigma^{t\bar{b}}_{SM})/\sigma^{t\bar{b}}_{SM}$,
as a function of $M_{\Pi_t}$, here the $\bar{t}$ production is
also considered as in the $tq$ production. The relative correction
drops sharply with $M_{\Pi_t}$. It is shown that even with large
$M_{\Pi_t}$, the contribution of the TC2 model can also enhance
the SM cross section significantly. To represent each contribution
of the TC2 model, we show
$\sigma^{t\bar{b}}_{SM},~\delta\sigma^{t\bar{b}}_{Wtb},~\sigma^{t\bar{b}}_{FC},~\sigma^{t\bar{b}}_{total}$
in Table.2. Similar to the $tq$ production, the correction induced
by the $Wtb$ coupling is negative and below $4\%$ in the parameter
space considered. The total contribution is dominated by the
contribution induced by the tree-level FC coupling
$\Pi_t^+c\bar{b}$.

Like the $tq$ production, there is the final state signature
$l\nu$ and a b-quark jet in the $t\bar{b}$ production. The other
energetic jet is also from a b-quark, and shares similar
kinematics with the b-quark from the t-quark decay. Thus b-quark
identification, or b tagging, in the $t\bar{b}$ production is
equally likely between the b-quark from the t-quark decay and the
b-quark from the original interaction. From a phenomenological
standpoint, the most important distinction of the final states
between $t\bar{b}$ and $tq$ productions is the presence of a
second high-$p_T$ b-jet in the $t\bar{b}$ process. In the $tq$
production, the second b-jet tends to be at low $p_T$ and is often
not seen. Therefore, the requirement of two b-jets with high $p_T$
will eliminate most of the background coming from the $tq$
production. On the other hand, the requirement of two b-tagged
jets is also crucial to reduce the contamination of W+jets events
that have a cross section several orders of magnitude that of the
signal. Furthermore, it is also necessary, as in other single
t-quark production modes, to design cuts to reduce the W+jets and
$t\bar{t}$ backgrounds. In order to reduce contamination by W+jets
events, the reconstructed t-quark mass in each event must fall
within a window about the known t-quark mass, and the events must
have a total transverse jet momentum above 175 GeV. Only events
containing exactly two jets(both tagged as b's) are kept in order
to reduce the $t\bar{t}$ background. The study indicates that,
despite the large anticipated background rate, it should be
possible to perform a good statistical measurement of the cross
section for the $t\bar{b}$ production. The resulting $S/B$ ratio
is about $11\%(9\%)$ in the $t\bar{b}(\bar{t}b)$ final state. It
is obvious that the combination of both final states is required
to improve the sensitivity. The precision on the cross section has
been assessed at the LHC for an integrated luminosity of $30
fb^{-1}$ at different stages of the analysis. After the simple
preselection stage, results show a good statistical sensitivity
but higher level of systematic uncertainties. The precision at
ATLAS for $L=30 fb^{-1}$ is shown as\cite{ATLAS precision}
\begin{eqnarray}
\frac{\Delta \sigma}{\sigma}=7\%_{stat.}\pm 13.8\%_{exp}\pm
11\%_{bckgd theo}\pm 5\%_{lumi}.
 \end{eqnarray}
 Using both the $H_T$(the total transverse energy) and reconstructed t-quark mass results in a
 significantly reduced level of systematics at the price of loss
 in statistical sensitivity
\begin{eqnarray}
\frac{\Delta \sigma}{\sigma}=12\%_{stat}\pm 12\%_{exp}\pm
11\%_{bckgd theo}\pm 5\%_{lumi}.
 \end{eqnarray}
 At CMS for $L=10 fb^{-1}$, the
precision is \cite{CMS precision}
 \begin{eqnarray}
\frac{\Delta \sigma}{\sigma}=18\%_{stat}\pm 31\%_{exp}\pm
3\%_{lumi}.
 \end{eqnarray}
In all case, systematic errors are expected to dominate the cross
section determination.

As we have discussed above, the measurement of the $t\bar{b}$
production may appear as the most delicate of the main three
single t-quark processes because of its relative low cross section
compared to the others at the LHC. It is however one of the most
interesting because the final state events of the $t\bar{b}$
production is directly sensitive to contributions from extra
particles predicted in new physics models. Our results shows that
the correction to the $t\bar{b}$ production induced by the $Wtb$
coupling is still small as in the $tq$ production which is
embedded in the large contribution coming from the tree-level FC
coupling in the TC2 model. Certainly, the LHC has the ability to
detect such FC effect. On the other hand, the LHC experiments
about the $t\bar{b}$ production should provide a severe limit on
the FC coupling $\Pi_t^+c\bar{b}$ if the cross section of the
$t\bar{b}$ production is measured precisely at the LHC. Therefore
$t\bar{b}$ production provides a unique chance to study the
properties of the FC coupling $\Pi_t^+c\bar{b}$ at the LHC.

Although the Tevatron can not measure three single t-quark
production modes separately, the measured cross section
$\sigma(p\bar{p}\rightarrow tb+X,tqb+X)$ has been given by the
Tevatron experiments which can also provide rude information about
$t\bar{b}$ production. The experimental value of the cross section
of $t\bar{b}$ production is almost consistent with the SM value.
So it is necessary to calculate the contribution of the TC2 model
to the $t\bar{b}$ production at the Tevatron due to the large
contribution of the TC2 model to the $t\bar{b}$ production. We
plot the total relative correction of the TC2 model to the
$t\bar{b}$ production at the Tevatron in Fig.8 and represent each
contribution in Table 3. Clearly there should exists large
down-limit on the top-pion mass if the predicted cross section in
the TC2 model is consistent with the Tevatron experiments.

\subsection{The associated production $tW$}

A single t-quark may also be produced via the weak interaction in
association with a real W boson $(q^2=M^2_W)$, as shown in
Fig.2(c)\cite{tw cross section-1,tw cross section-2}. Like the
$tq$ production, one of the initial partons is a b-quark. However,
unlike the $tq$ production, this associated production scales like
1/s. This, combined with the higher values of x needed to produce
both a t-quark and a W boson, leads to a cross section which is
significantly less than that of the $tq$ process, despite the fact
that it is order $\alpha_s\alpha_e$ rather than $\alpha^2_e$. The
$tW$ process is also known at NLO, and the total NLO cross section
is 0.14 pb at the Tevatron which is negligible, while it is 68 pb
at the LHC\cite{t-theoretical uncertainty-3,tw cross section-2,tw
NLO}.

If the t-quark has indeed a special role in the generation of
masses, it is crucial that its interactions should be carefully
studied in order to learn what properties the underlying theory at
high energies must possess. The deviations of  $Wtb$ coupling from
the SM predictions may represent the best clues on the nature of
the EWSB. As we know, each single t-quark production mode is
sensitive to different types of new physics, with the $tW$ mode
distinct in that it is sensitive only to physics which directly
modifies the $Wtb$ coupling from its SM structure. This
distinction is a result of the fact that in this mode both the
t-quark and the W are directly observable, whereas in the other
two modes the W bosons are virtual, and thus those processes may
receive contributions from exotic types of charged bosons or FCNC
interactions. On the other hand, $tW$ mode can also provide
complimentary information about the $Wtb$ coupling by probing it
in a region of momentum different from other single t-quark
production modes. Similar to the $tq$ production, the cross
section for this associated production increases by more than two
orders of magnitude from the Tevatron to the LHC, it is
sufficiently large at the LHC to not only observe this mode of
single t-quark production but also to study the $Wtb$ coupling in
detail.

Unlike the other two single t-quark production modes, there only
exists the contribution from $Wtb$ coupling for the $tW$
production in the TC2 model, shown as Fig.9. The production
amplitudes including such contribution are
\begin{eqnarray}
 M_{8}=-g_{s}T^{a}_{ij}\overline{u}_{ti}(p_{4})\Gamma^{\mu}(p_4, p_3+p_4)\varepsilon_{\mu}(p_{3})\frac{\pslash_3+\pslash_4+m_{b}}
 {(p_{3}+p_{4})^{2}-m^{2}_{b}}\not\varepsilon^{a}(p_{1})u_{bj}(p_{2}),\\
 M_{9}=-g_{s}T^{a}_{ij}\overline{u}_{ti}(p_{4})\not\varepsilon^{a}(p_{1})\frac{\pslash_2-\pslash_3+m_{t}}
 {(p_{2}-p_{3})^{2}-m^{2}_{t}}\varepsilon_{\mu}(p_{3})\Gamma^{\mu}(p_2-p_3,p_2)u_{bj}(p_{2}).
 \end{eqnarray}
The total cross section of the $tW$ production is defined as
$\sigma^{tW}_{total}=\sigma^{tW}_{SM}+\delta\sigma^{tW}_{Wtb}$.
Here we only consider the tree-level SM cross section
$\sigma^{tW}_{SM}$, and $\delta\sigma^{tW}_{Wtb}$ represents the
correction from the $Wtb$ coupling. In the calculation, we take
the same parameter values as in the other two single t-quark
production modes, and also consider the $\bar{t}$ production. Here
we focus on studying the $tW$ production mode at the LHC. The
numerical results of relative correction at the LHC,
$(\sigma^{tW}_{total}-\sigma^{tW}_{SM})/\sigma^{tW}_{SM}$, are
shown in Fig.10. Because the correction to the $tW$ production
only comes from the modification of the $Wtb$ coupling in the TC2
model the relative correction to the $tW$ production is small and
negative.

As for the other two single t-quark production modes, we select
$tW$ events by requiring a single high $P_T$ lepton and a high
missing transverse energy. Such a selection criterion implies that
one W boson decays leptonically and that the second W boson must
decay into two jets. Therefore, the selected events have exactly
three jets with one of them tagged as a b-jet. This allows to
reject part of $t\bar{t}$ background. In addition, by requiring a
2-jet invariant mass within a window around the W mass, it is
possible to eliminate most events that do not contain a second W,
i.e. all backgrounds other than $t\bar{t}$.

The strategy for measuring the cross section of the $tW$
production is similar to that for the $tq$ production, as they
share the same backgrounds. However, the nature of the associated
production makes it relatively easy to separate the signal from
$W+$jets background and difficult to separate from $t\bar{t}$
background. Therefore, the dominant background arises from
$t\bar{t}$ production. From the point of view of signal
identification, the region with small $tW$ invariant mass, near
the $t\bar{t}$ threshold, has possibly a chance to be optimal. The
$tW$ process analysis benefits from the relative high cross
section. However, due to high similarities with top pair events,
the selection is hampered by a high level of background
contamination. This characteristics makes the $tW$ cross section
very difficult to measure with the early data at the LHC. Two
studies designed to separate signal from background have been
performed using two different final states. The first is a study
which attempts to isolate $tW$ signal events in which one W decays
to jets and the other decays to leptons\cite{study1}. The second
study attempts to isolate signal events in which both W's decay
leptonically\cite{study2}. Based on the SM prediction, the $S/B$
ratio is well below $10\%$. Because the main background comes from
the top pair production, the prior precise determination of the
top pair production cross section is needed. Combining both
electron and muon channels as well as all two and three jet final
states leads to a statistical precision slightly below $6\%$ for
an integrated luminosity of $1 fb^{-1}$. As we have discussed
above, the relative correction of the TC2 model to the $tW$
production is only a few percent even in the optimal parameter
space. Based on the precision on the cross section of the $tW$
production, it is difficult to observe such correction at the LHC.
Therefore, $tW$ production is not ideal process to test the TC2
model.

\section{The summary and conclusion}

The single t-quark production plays an important role in probing
the properties of the $Wtb$ coupling and new physics model. With
the running of the LHC, it is possible to measure each single
t-quark production mode separately and the cross section can be
measured precisely. So the LHC opens an ideal window to probe the
new physics via the single t-quark production. In this paper, we
systematically study the contribution of the TC2 model to each
single t-quark production mode at the LHC and also study the
$t\bar{b}$ at the Tevatron. Due to the existence of the top-pions
and top-higgs, the TC2 model can contribute to the single t-quark
production in two different ways. One is that the TC2 model can
make the loop-level contribution to the cross sections which is
induced by the modification of the $Wtb$ coupling in the TC2
model. Such contribution exists in all three single t-quark
production modes, but it is very small. Even at the LHC, it is
difficult to observe this kind of loop-level contribution. Another
kind of contribution comes from the tree-level FC couplings in the
TC2 model. Such contribution only exists in the $tq$ and
$t\bar{b}$ production modes and the resonant effect can greatly
enhance the cross sections. The recent Tevatron experiments has
given a cross section for single t-quark production of
$\sigma(p\bar{p}\rightarrow tb+X,tqb+X)$ which is consistent with
the SM value. So there should exists down-limit on the top-pion
mass. With the precise measurement of the cross section for each
single t-quark production mode at the LHC, detailed information
about the parameters and the FC couplings in the TC2 model can be
obtained.

\section{Acknowledgments}
\hspace{1mm}

This work is supported  by the National Natural Science Foundation
of China under Grant No.10775039, 10575029 and 10505007.

\newpage

\begin{center}
{\bf Table 1: Each contribution of the TC2 model to the $tq$ production at the LHC}\\
\vspace{1cm}
 \doublerulesep 0.8pt \tabcolsep 0.05in
\begin{tabular}{|c|c|c|c|c|c|c|}\hline \hline
$\varepsilon$ &$M_{h_{t}}(GeV)$ &$M_{\Pi_t}(GeV)$ &
$\sigma^{tq}_{SM} (pb)$& $\delta\sigma^{tq}_{Wtb} (pb)$&
$\sigma^{tq}_{FC}(pb)$ & $\sigma^{tq}_{total}
(pb)$\\
\cline{4-7}
 \hline
    &    &200 &           &-7.24          &23.26  &239.42            \\
    &    &250 &           &-7.65          &14.44  &230.19           \\
0.03&200 &300 &223.40     &-8.21          &9.87   &225.06  \\
    &    &350 &           &-8.83          &7.12   &221.69           \\
    &    &400 &           &-9.47          &5.64   &219.56       \\
 \cline{2-7}
\hline
    &    &200 &           &-6.80          &38.64  &255.24            \\
    &    &250 &           &-7.19          &23.78  &239.98            \\
0.06    &200 &300 &223.40 &-7.71           &15.78  &231.46      \\
    &    &350 &           &-8.30          &11.22  &226.33             \\
    &    &400 &           &-8.90         &8.53   &223.03            \\
 \cline{2-7}
    \hline
    &    &200     &        &-6.24          &55.16  &272.32             \\
    &    &250    &          &-6.60         &33.45  &250.25            \\
0.1    &200 &300  &223.40  &-7.08    &21.87  &238.19      \\
    &    &350    &         &-7.61          &15.18  &230.97            \\
    &    &400    &         &-8.17          &11.32  &226.55          \\
 \cline{2-7}
 \hline\hline
\end{tabular}
\end{center}

\newpage
\begin{center}
{\bf Table 2: Each contribution of the TC2 model to the $t\bar{b}$
production at the LHC}\\
 \vspace{1cm}
 \doublerulesep 0.8pt \tabcolsep 0.05in
\begin{tabular}{|c|c|c|c|c|c|c|}\hline \hline
$\varepsilon$ &$M_{h_{t}}(GeV)$ &$M_{\Pi_t}(GeV)$ &
$\sigma^{t\bar{b}}_{SM} (pb)$& $\delta\sigma^{t\bar{b}}_{Wtb}
(pb)$& $\sigma^{t\bar{b}}_{FC}(pb)$ & $\sigma^{t\bar{b}}_{total}
(pb)$\\
\cline{4-7}
   \hline
  &          &200          &       &-0.07    &1315.66   & 1322.81\\
  &          &250          &       &-0.14    &1827.92   & 1834.99 \\
0.03   &200  &300          &7.21   &-0.19    &1261.03   & 1268.05  \\
    &        &350          &       &-0.24    &789.16     & 796.14 \\
    &        &400          &       &-0.28    &492.08     & 499.01 \\
 \cline{2-7}
  \hline
    &        &200          &      &-0.06      &1289.54    & 1296.68 \\
    &        &250          &      &-0.13      &1582.34    & 1589.42 \\
0.06    &200 &300          &7.21  &-0.18      &1087.66     & 1094.68 \\
    &        &350          &      &-0.22      &683.81      &  690.79\\
    &        &400          &      &-0.26      &430.73      &  437.68\\
 \cline{2-7}
    \hline
    &        &200          &      &-0.06       &1231.46    &  1238.61 \\
    &        &250          &      &-0.12       &1341.44    &  1348.53 \\
0.1    &200  &300          &7.21  &-0.17       &903.13     & 910.17 \\
    &        &350          &      &-0.21       &567.59     &  574.60 \\
    &        &400          &      &-0.24       &359.65     &   366.62\\
 \cline{2-7}
 \hline\hline
\end{tabular}
\end{center}

\newpage
\begin{center}
{\bf Table 3: Each contribution of the TC2 model to the $t\bar{b}$
production at the Tevatron}\\
 \vspace{1cm}
 \doublerulesep 0.8pt \tabcolsep 0.05in
\begin{tabular}{|c|c|c|c|c|c|c|}\hline \hline
$\varepsilon$ &$M_{h_{t}}(GeV)$ &$M_{\Pi_t}(GeV)$ &
$\sigma^{t\bar{b}}_{SM} (pb)$& $\delta\sigma^{t\bar{b}}_{Wtb}
(pb)$& $\sigma^{t\bar{b}}_{FC}(pb)$ & $\sigma^{t\bar{b}}_{total}
(pb)$\\
   \hline
  &          &200          &       &  -0.004  & 5.61 &5.86 \\
  &          &250          &       &  -0.007  & 4.98 & 5.23\\
0.03   &200  &300          & 0.26  &  -0.008 & 2.15 & 2.40 \\
    &        &350          &       &  -0.010 &0.86    & 1.10\\
    &        &400          &       &  -0.010 & 0.35   & 0.59\\
  \hline
    &        &200          &      &  -0.004    &  5.00  &5.24  \\
    &        &250          &      &  -0.006   & 4.16  & 4.41 \\
0.06    &200 &300          &  0.26& -0.008   &  1.85 & 2.10 \\
    &        &350          &      &  -0.009   & 0.77    & 1.02 \\
    &        &400          &      & -0.010   &0.34    & 0.58 \\
    \hline
    &        &200          &      & -0.004     & 4.34  &  4.60 \\
    &        &250          &      &   -0.006   & 3.32  &  3.57 \\
0.1    &200  &300          &  0.26&    -0.007  & 1.51  &1.76 \\
    &        &350          &      &   -0.008   & 0.66 & 0.91  \\
    &        &400          &      &  -0.009    &  0.30  &  0.55 \\
 \hline\hline
\end{tabular}
\end{center}

\newpage
\begin{figure}[h]
\begin{center}
\epsfig{file=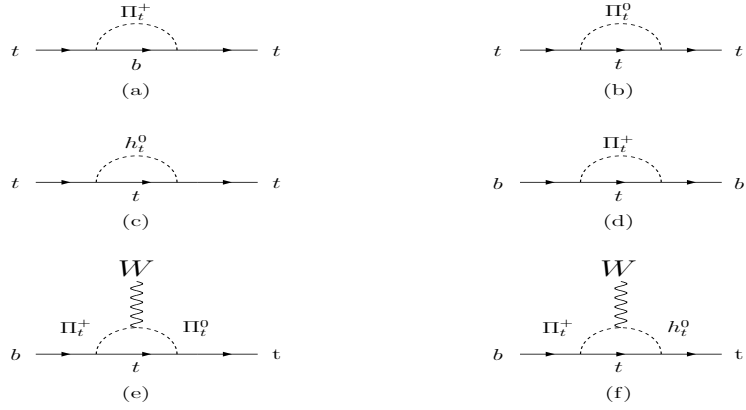,width=450pt,height=420pt} \vspace{-7.5cm}
\caption{\small  The one-loop contribution of the TC2 model to the
$Wtb$ coupling.}
\end{center}
\end{figure}

\begin{figure}[h]
\begin{center}
\epsfig{file=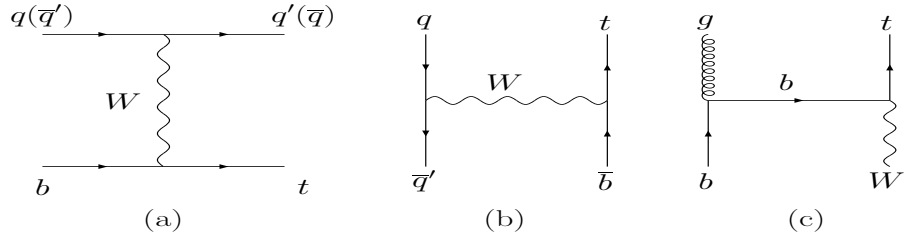,width=450pt,height=420pt} \vspace{-10.5cm}
\caption{\small  The leading-order Feynman diagrams for the single
t-quark production in the SM. (a)For the $tq$ process, (b)for the
$t\bar{b}$ process, and(c) for the associated $tW$ process.}
\end{center}
\end{figure}

\newpage

\begin{figure}[h]
\begin{center}
\epsfig{file=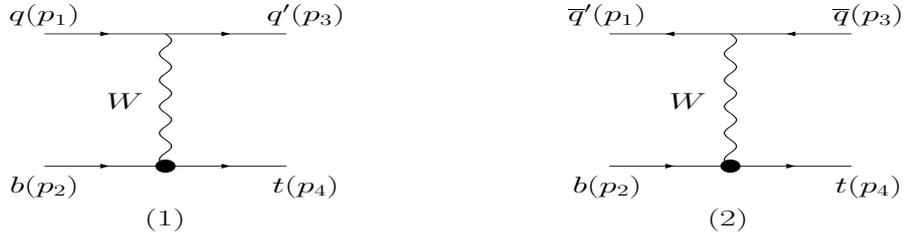,width=450pt,height=420pt} \vspace{-10.5cm}
\caption{\small The contribution to the $tq$ production induced by
the modification of the $Wtb$ coupling in the TC2 model.}
\end{center}
\end{figure}

\begin{figure}[h]
\begin{center}
\epsfig{file=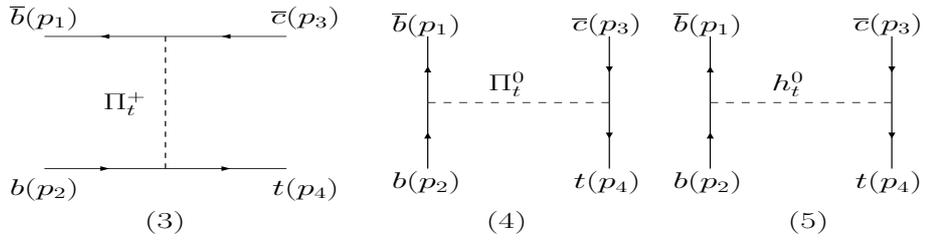,width=450pt,height=420pt} \vspace{-10.5cm}
\caption{\small The tree-level contribution of the FC couplings in
the TC2 model to the $t\bar{c}$ production.}
\end{center}
\end{figure}

\newpage

\begin{figure}[h]
\scalebox{0.8}{\epsfig{file=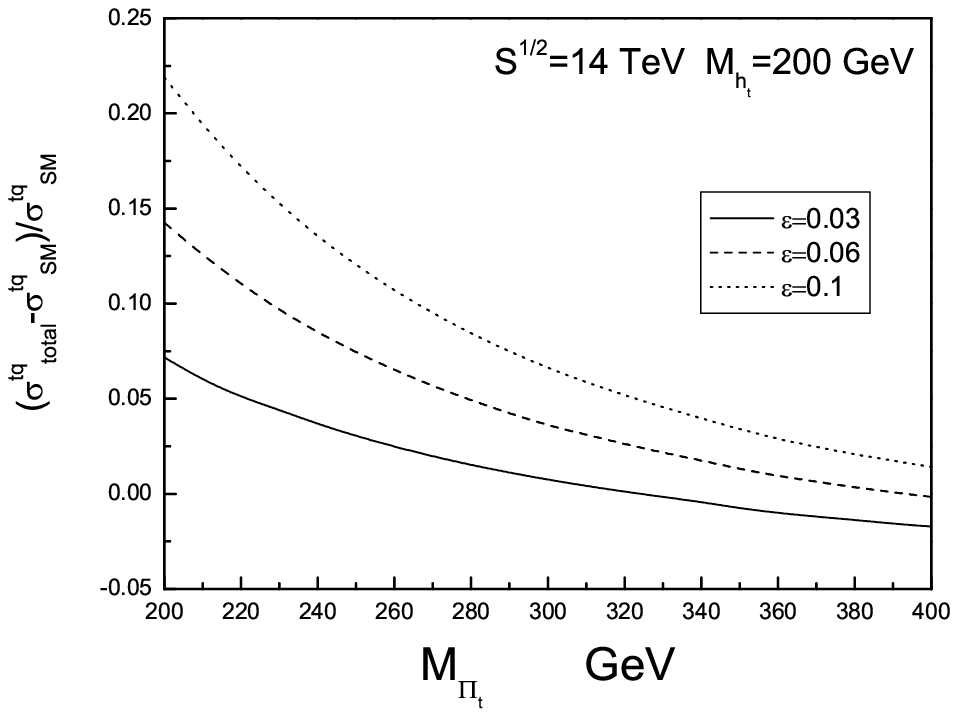}}
\scalebox{0.8}{\epsfig{file=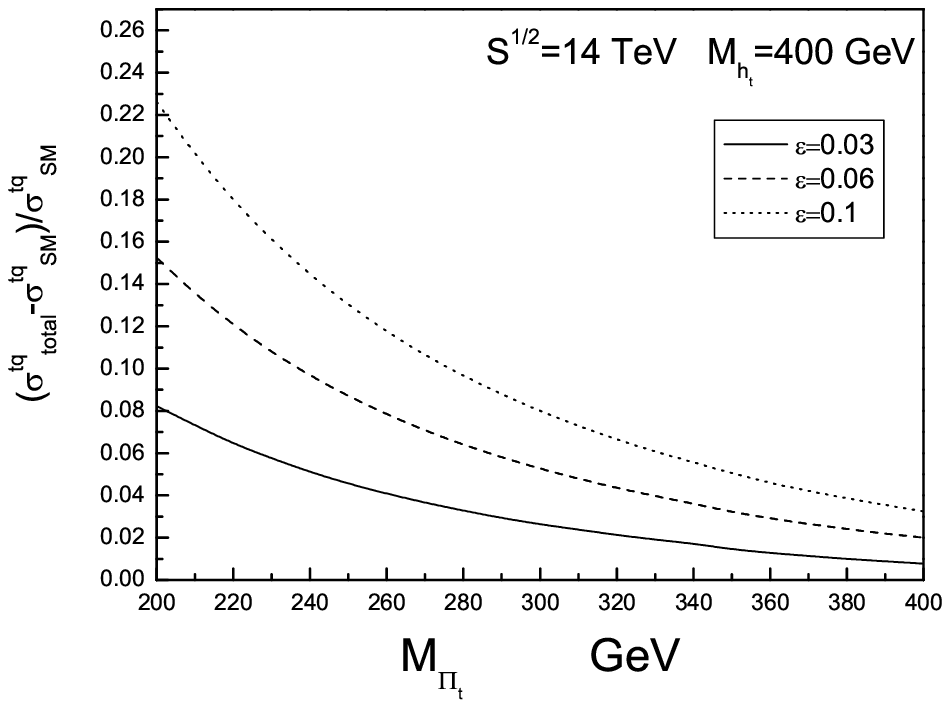}}\\
\caption{\small The relative correction of the TC2 model to the
$tq$ production at the LHC,
$(\sigma^{tq}_{total}-\sigma^{tq}_{SM})/\sigma^{tq}_{SM}$, as a
function of $M_{\Pi_t}$ with $\varepsilon$ being 0.03,0.06,0.1 and
$M_{h_t}$ being 200, 400 GeV.}
\end{figure}

\begin{figure}[h]
\begin{center}
\epsfig{file=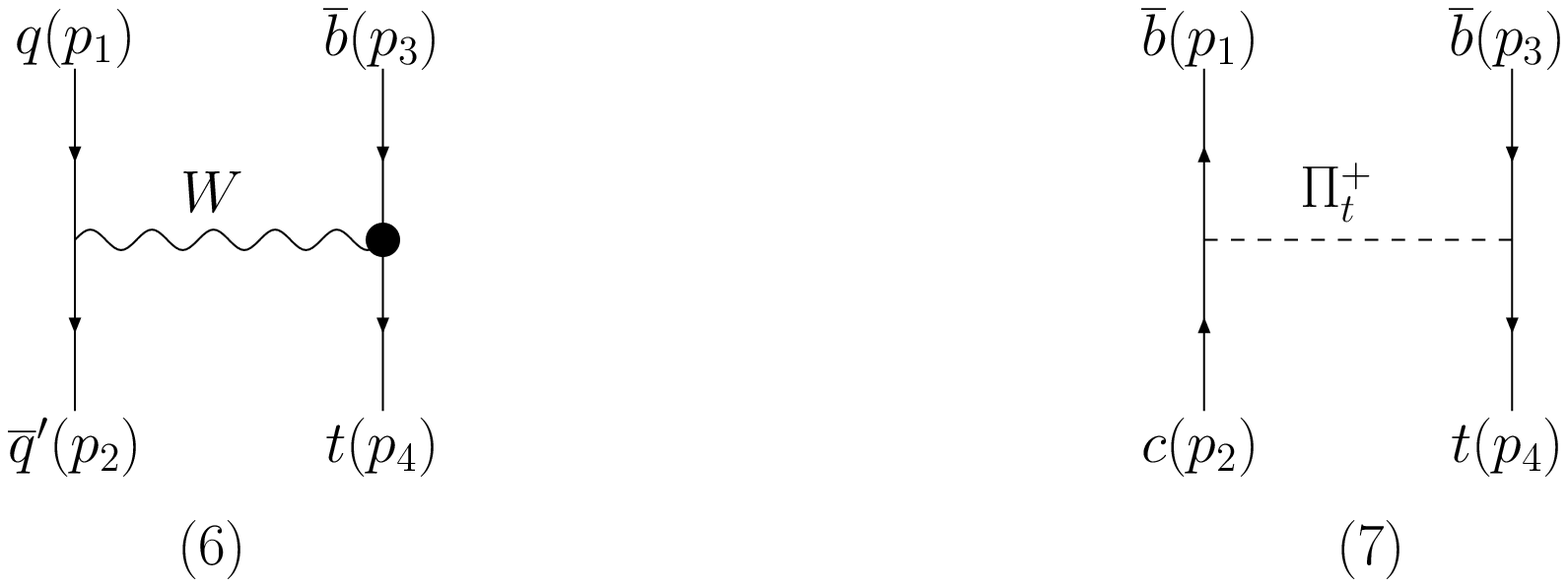,width=450pt,height=420pt} \vspace{-10.5cm}
\caption{\small The contribution of the TC2 model to the
$t\bar{b}$ production.}
\end{center}
\end{figure}

\newpage

\begin{figure}[h]
\scalebox{0.8}{\epsfig{file=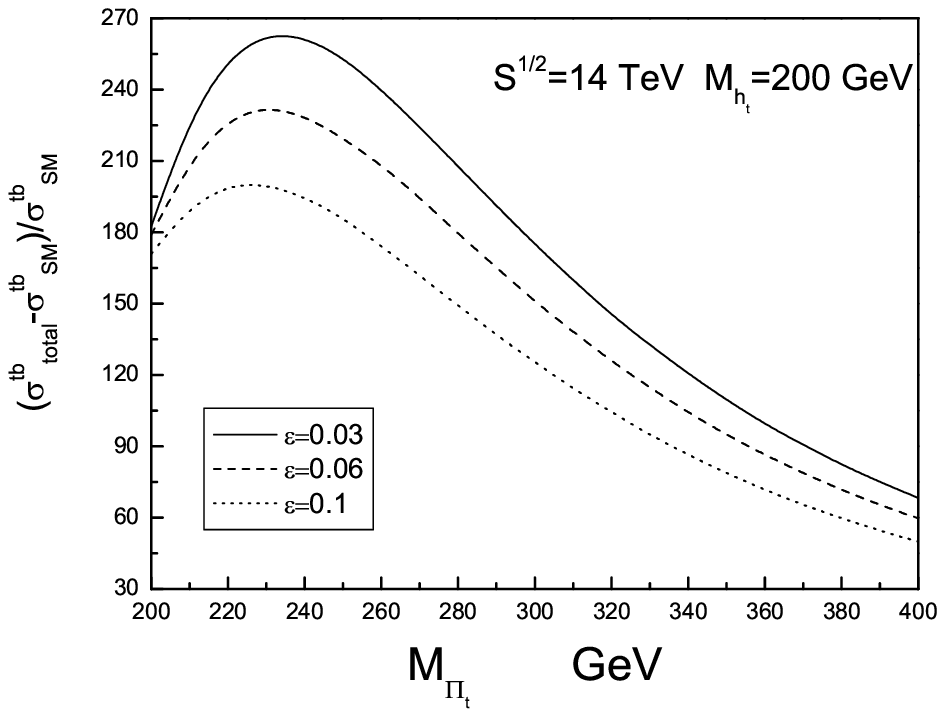}}
\scalebox{0.8}{\epsfig{file=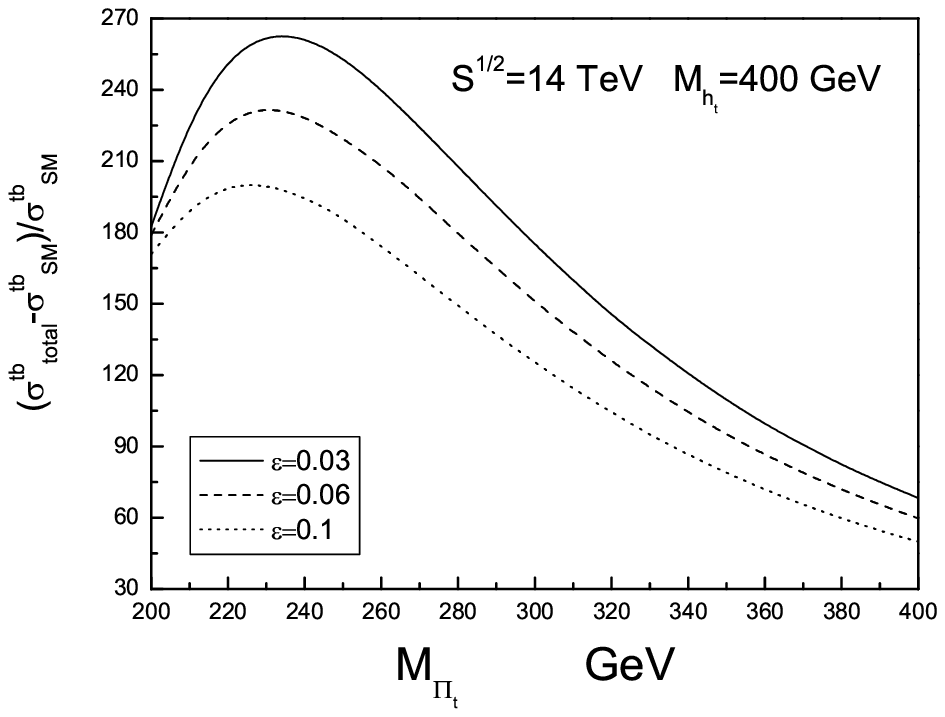}}\\
\caption{\small The relative correction of the TC2 model to the
$t\bar{b}$ production at the LHC,
$(\sigma^{t\bar{b}}_{total}-\sigma^{t\bar{b}}_{SM})/\sigma^{t\bar{b}}_{SM}$,
as a function of $M_{\Pi_t}$ with $\varepsilon$ being
0.03,0.06,0.1 and $M_{h_t}$ being 200, 400 GeV.}
\end{figure}

\begin{figure}[h]
\scalebox{0.8}{\epsfig{file=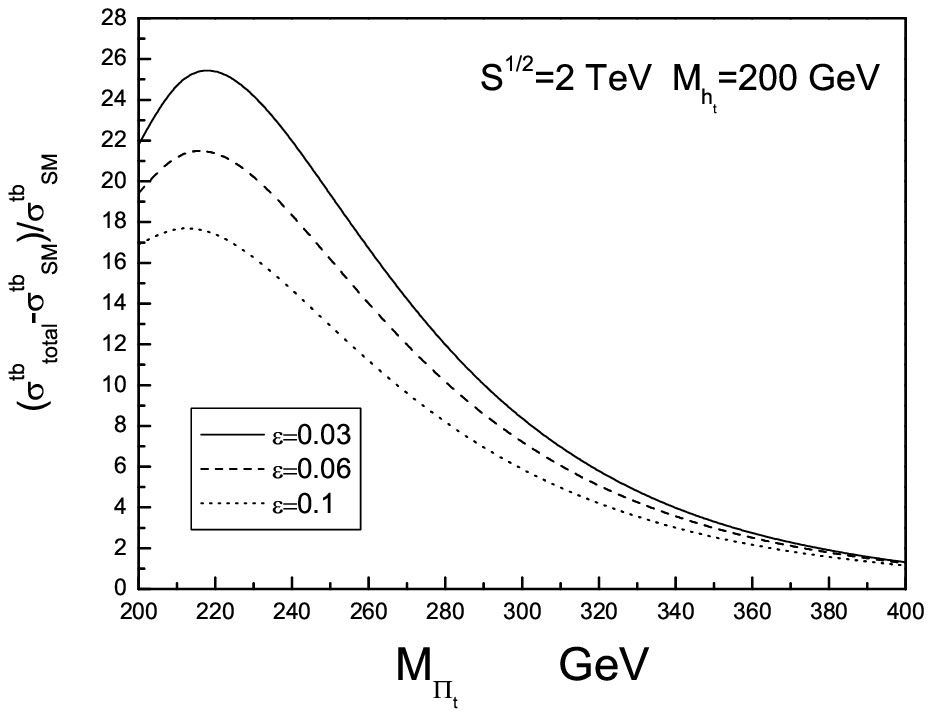}}
\scalebox{0.8}{\epsfig{file=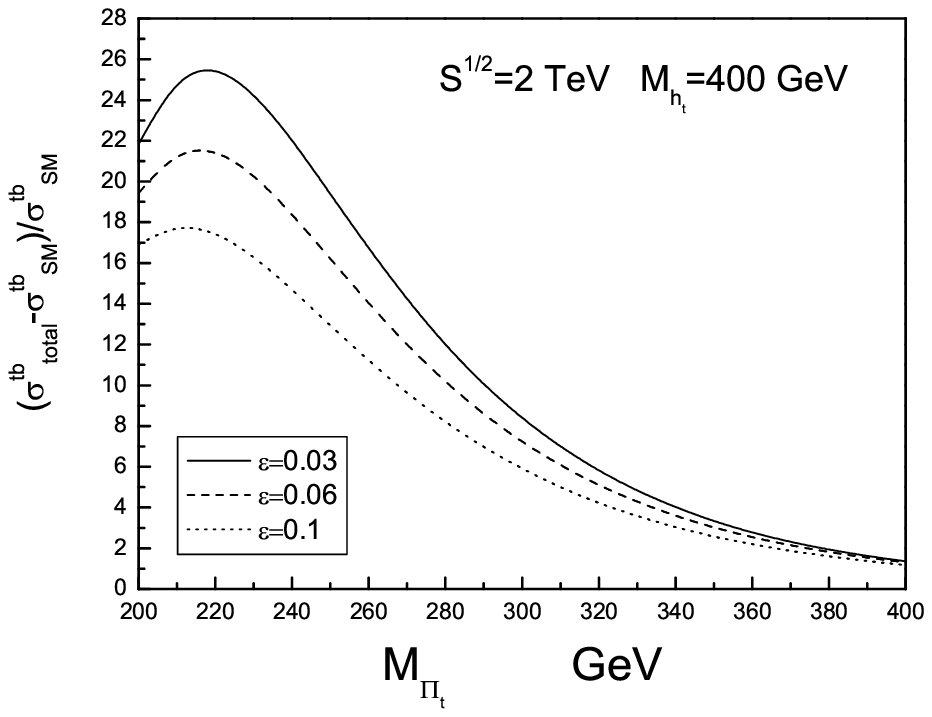}}\\
\caption{\small The relative correction of the TC2 model to the
$t\bar{b}$ production at the Tevatron,
$(\sigma^{t\bar{b}}_{total}-\sigma^{t\bar{b}}_{SM})/\sigma^{t\bar{b}}_{SM}$,
as a function of $M_{\Pi_t}$ with $\varepsilon$ being
0.03,0.06,0.1 and $M_{h_t}$ being 200, 400 GeV.}
\end{figure}

\newpage

\begin{figure}[h]
\begin{center}
\epsfig{file=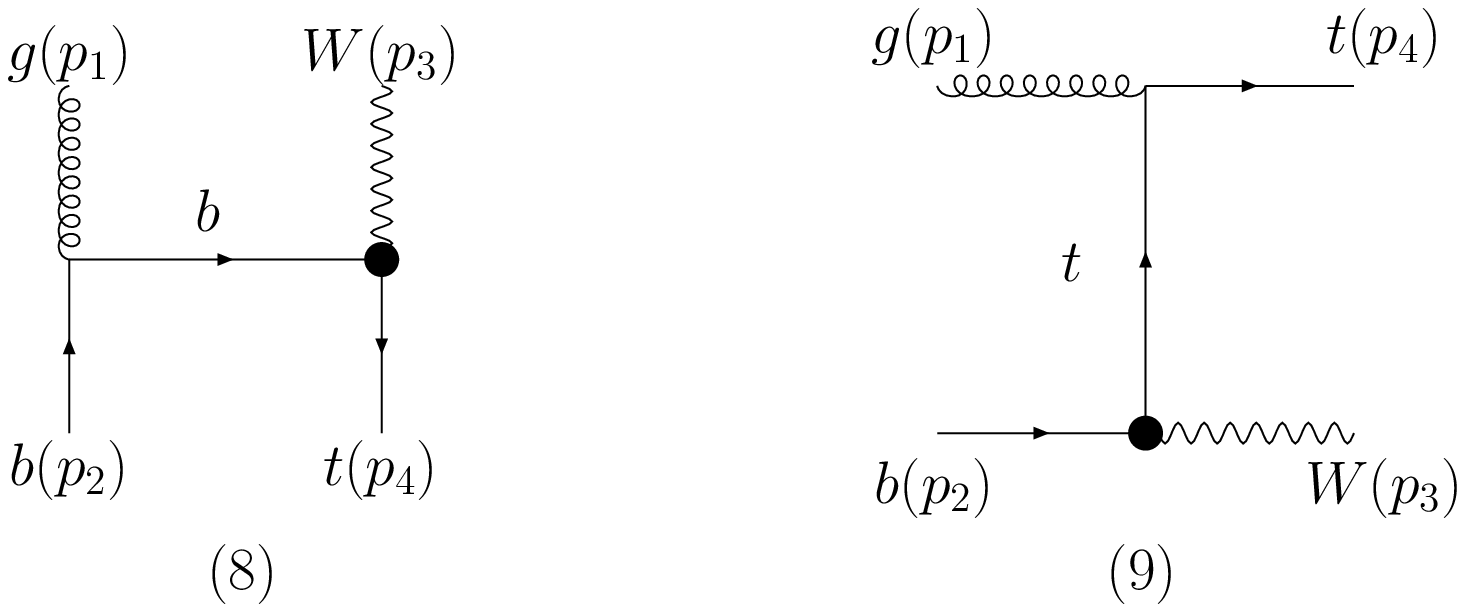,width=450pt,height=450pt} \vspace{-10.5cm}
\caption{\small The contribution to the associated $tW$ production
induced by the modification of the $Wtb$ coupling in the TC2
model.}
\end{center}
\end{figure}

\begin{figure}[h]
\scalebox{0.8}{\epsfig{file=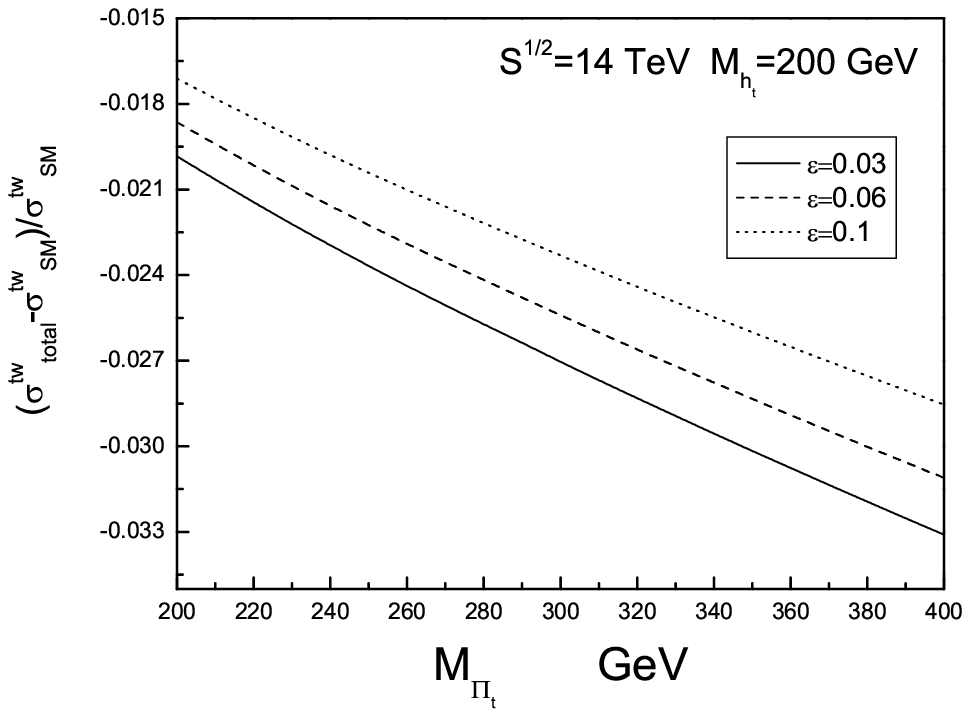}}
\scalebox{0.8}{\epsfig{file=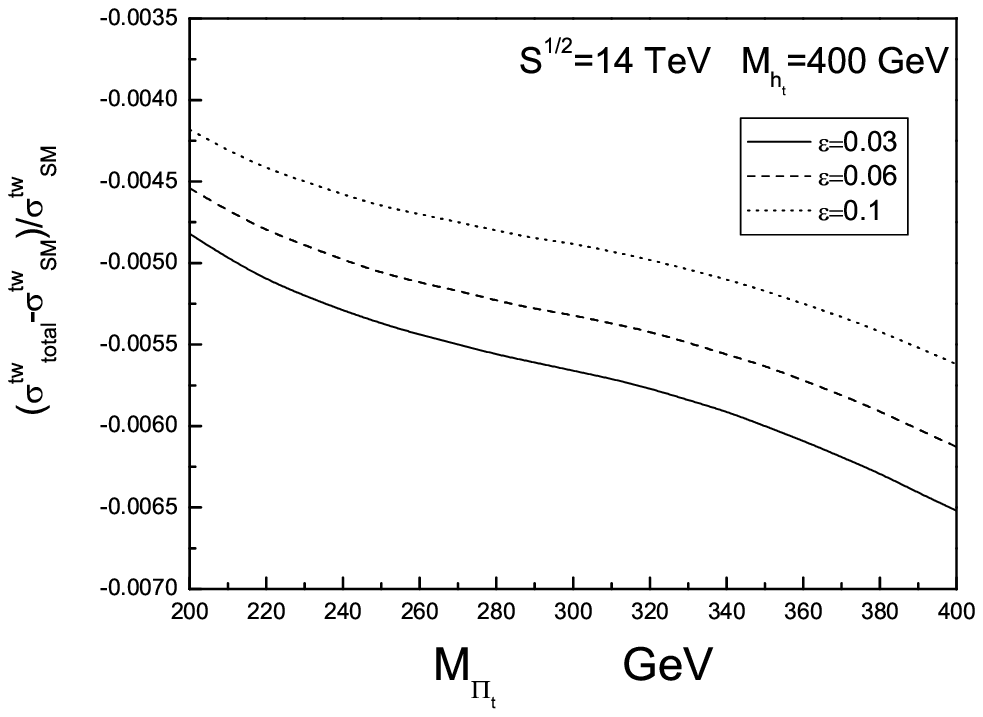}}\\
\caption{\small The relative correction of the TC2 model to the
$tW$ production at the LHC,
$(\sigma^{tW}_{total}-\sigma^{tW}_{SM})/\sigma^{tW}_{SM}$, as a
function of $M_{\Pi_t}$ with $\varepsilon$ being 0.03,0.06,0.1 and
$M_{h_t}$ being 200, 400 GeV.}
\end{figure}

\end{document}